\documentclass[10pt,journal,compsoc]{IEEEtran}
\newcommand{\subparagraph}{}
\usepackage{titlesec}
\titlespacing{\section}{0pc}{5pt}{2pt}
\titlespacing{\subsection}{0pc}{4pt}{1pt}

\usepackage{ifpdf}

\usepackage{enumitem}

\ifCLASSOPTIONcompsoc
  \usepackage[nocompress]{cite}
\else
  \usepackage{cite}
\fi

\ifCLASSINFOpdf
\else
\fi

\usepackage{algorithmic}

\ifCLASSOPTIONcompsoc
 \usepackage[caption=false,font=footnotesize,labelfont=sf,textfont=sf]{subfig}
\else
 \usepackage[caption=false,font=footnotesize]{subfig}
\fi

\usepackage{url}

\usepackage{float}
\usepackage{amsmath,amssymb,amsfonts}
\usepackage{graphicx}
\usepackage{textcomp}
\usepackage{multirow}
\usepackage[dvipsnames]{xcolor}
\usepackage{colortbl}
\usepackage{inputenc}
\usepackage{diagbox}
\usepackage{mathrsfs}
\usepackage{threeparttable}
\usepackage{listings}
\usepackage{booktabs}
\usepackage{hhline}
\usepackage{array,ragged2e}
\usepackage{tabularx}
\usepackage{fancyhdr} 
\usepackage{makecell}
\usepackage{courier}
\usepackage{setspace}

\usepackage[font={footnotesize,sf}]{caption}

\begin{document}
%
\title{\huge Architectural Implications of Graph Neural Networks}
%
%
%
%

\author{~Zhihui~Zhang$^{1\star}$,~Jingwen~Leng$^{1\star\circ}$,~Lingxiao~Ma$^2$,~Youshan~Miao$^3$,~Chao~Li,$^1$~Minyi~Guo$^{1\circ}$\\%
\textit{\footnotesize$\lbrace$zhihui.zhang, leng-jw$\rbrace$@sjtu.edu.cn, xysmlx@pku.edu.cn, yomia@microsoft.com, $\lbrace$lichao, guo-my$\rbrace$@cs.sjtu.edu.cn}
\IEEEcompsocitemizethanks{
\IEEEcompsocthanksitem $^1$Shanghai Jiao Tong University, $^2$Peking University, $^3$Microsoft Research. $^{\circ}$Corresponding authors: Jingwen Leng and Minyi Guo.
\IEEEcompsocthanksitem $^\star$The work is done when Zhihui Zhang is an intern and Jingwen Leng is a visiting researcher at Microsoft Research.
}}

%
%

\markboth{IEEE COMPUTER ARCHITECTURE LETTERS,~Vol.~19, No.~1, JANUARY-JUNE~2020}%
{Zhihui \MakeLowercase{\textit{et al.}}: Analyzing the Computations in Graph Neural Networks}
%


\IEEEtitleabstractindextext{%
\begin{abstract}
Graph neural networks (GNN) represent an emerging line of deep learning models that operate on graph structures. 
It is becoming more and more popular due to its high accuracy achieved in many graph-related tasks.
However, GNN is not as well understood in the system and architecture community as its counterparts such as multi-layer perceptrons and convolutional neural networks.
This work tries to introduce the GNN to our community. In contrast to prior work that only presents characterizations of GCNs, our work covers a large portion of the varieties for GNN workloads based on a general GNN description framework.
By constructing the models on top of two widely-used libraries, we characterize the GNN computation at inference stage concerning general-purpose and application-specific architectures and hope our work can foster more system and architecture research for GNNs.
\end{abstract}

}

\newcommand{\fixme}[1]{{\textcolor{red}{\textit{#1}}}}
\newcommand{\proj}{\textsc{SGBench\xspace}}
\newcommand{\Fig}[1]{Figure~\ref{#1}}
\newcommand{\Figsub}[1]{Figure~\subref{#1}}
\newcommand{\Tbl}[1]{Table~\ref{#1}}
\newcommand{\Sec}[1]{Section~\ref{#1}}
\newcommand\tab[1][16mm]{\hspace*{#1}}
\newcommand{\TODO}[1]{{\color{red} {\bf #1}}}
\newcommand{\pn}{{SGBench}}
\newcommand{\sagaann}{{SAGA-NN}}
\renewcommand{\paragraph}[1]{\noindent\textbf{#1}\hspace*{.05cm}}

\lstset{
    emph={SAGA,2,-,NN,Layer,ApplyEdge,ApplyVertex,Scatter,Gather,reture,Aggregate,(,)},
    emphstyle={\textbf}, 
    emph={[2]Accum,A[u]},
    emphstyle={[2]\color{black}},
    language={[]Java},
    basicstyle=\footnotesize\ttfamily, 
    keywordstyle=\color{blue}\bfseries, 
    commentstyle=\itshape\color{green!40!black}, frame=single
}

\graphicspath{ {./fig/} }

\maketitle

\IEEEdisplaynontitleabstractindextext

%
\IEEEpeerreviewmaketitle

\IEEEraisesectionheading{
\section{Introduction}\label{sec:introduction}}
 
\IEEEPARstart{G}{raph} neural networks (GNN) start to gain momentum as researchers are considering important tasks involving the \textit{graph} structure such as social media. 
Deep learning (DL) has achieved great success in domains with \textit{grid} data structure, e.g., images and sequences, which, however, only represents a small portion of the real-world data.
In contrast, \textit{graph} structure reflects the vast majority of real-world data such as molecular structure and knowledge graph.

Graph representation learning is one of the most important graph-related problems~\cite{Leskovec}.
It converts the irregular graph structure to embedding vectors, which are the compressed representations of vertices (i.e. vertex embedding) and the entire graph (i.e. graph embedding).
A downstream task such as molecular property prediction can then take in the regular embeddings rather than the raw graph for efficient processing.
As such, the quality of the embeddings directly determines the accuracy of downstream tasks.

Traditional graph representation methods like DeepWalk~\cite{DeepWalk} and node2vec~\cite{node2vec} mostly rely on hand-crafted or intuition-based algorithms.
In contrast, GNNs extend the graph analytics with DL's end-to-end learning capability, which has led to better accuracies in a variety of domains including molecular science, recommendation, and transportation.  
To realize the full potential of GNNs, we should adapt the existing software and hardware platforms to GNNs' unique characteristics. 

\begin{figure}[t]
    \centering
    \vspace*{-24pt}
    \includegraphics[trim=3.7cm 2.8cm 1.8cm 2.5cm,clip,width=\linewidth]{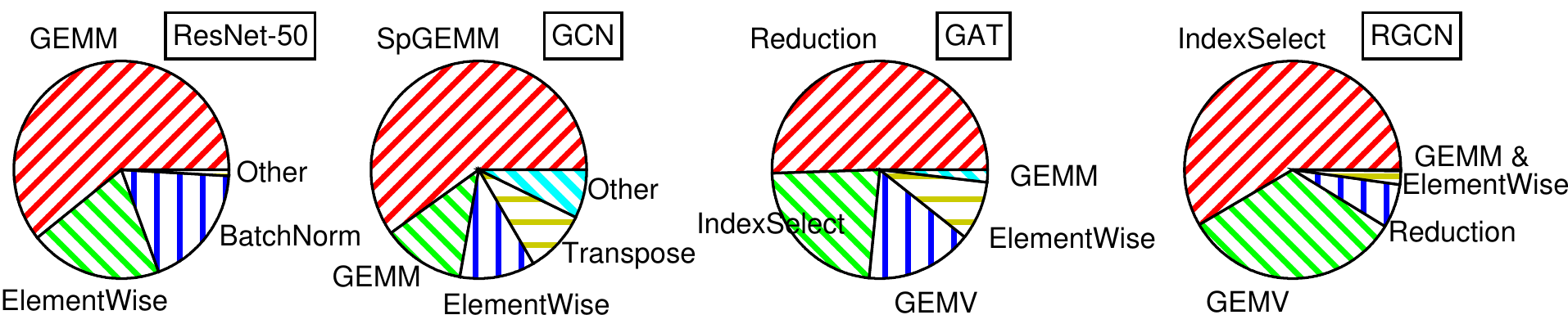}
    \vspace*{-17pt}
    \caption{Kernel breakdown of CNN (ResNet-50) and three GNN models.}
    \label{fig:cnn_vs_gnn_breakdowna}
    \vspace*{-18pt}
\end{figure}

The combination of DL and graph analytics makes GNN a new computation paradigm, which is quite different from their counterparts such as multi-layer perceptrons (MLP) and convolutional neural networks (CNN).
\Fig{fig:cnn_vs_gnn_breakdowna} compares the graphics processing unit (GPU) kernel distribution for ResNet-50 and three popular GNNs.
It is well understood that CNNs are dominated by convolutional layers, which are implemented through general matrix multiplication (GEMM) on GPU.
In contrast, the computation-intensive GEMM kernel is not the hotspot in the three GNNs, which also demonstrate model-specific patterns.

In this work, we aim to introduce GNN to our community. In contrast to prior work~\cite{cal20} that only presents characterizations of GCN~\cite{GCN_new}, we select five representative GNN models that cover a large portion of the varieties for GNN workloads on the basis of a general GNN description framework and our model review.
By constructing the models on top of two widely-used libraries, we characterize the efficiency of the GNN computation at inference stage on the existing GPU architecture and suggest directions for GNN-specific accelerators.
We hope that our analysis can help architects and system designers have a better understanding of GNN computation and foster more future work.

\begin{figure}[b]
    \centering
    \vspace{-18pt}
    \includegraphics*[width=0.95\linewidth]{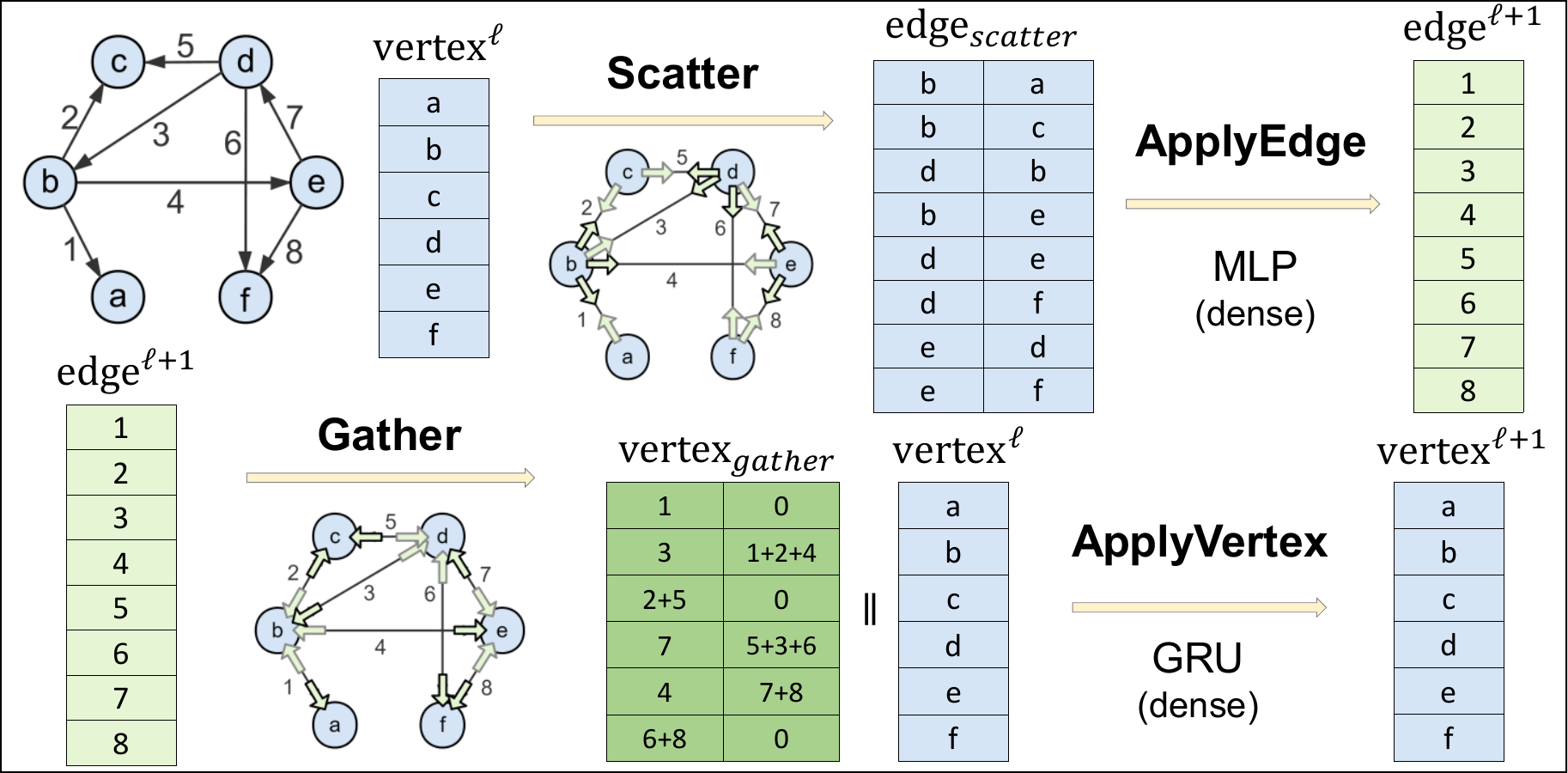}\\
    \vspace{-5pt}
    \caption{The computation stages in a GGNN~\cite{GG-NN} layer.}
    \label{fig:saga2-nn-stage}
    \vspace{-3pt}
\end{figure}

\section{GNN Benchmark Suite Construction}

This section describes our methodology of constructing a representative GNN benchmark suite.
We use a general description framework to perform a detailed review of the recently published GNN models.
The result identifies a few common patterns across the surveyed models, which lets us choose five models to cover almost all patterns.

\paragraph{Model Survey.}
A multi-layered GNN model is designed to learn the embeddings (i.e., vectors) for each vertex and edge in a graph.
The input for a GNN layer is the graph structure in the form of adjacency matrix, together with the vertex (edge) embedding matrix $vertex^l$ ($edge^l$). The layer generates the transformed embedding matrix $vertex^{l+1}$ and $edge^{l+1}$ for the next layer, as shown in \Fig{fig:saga2-nn-stage}.
We reviewed 53 GNN models published in recent top conferences and journals, spanning across molecular science, recommendation, and transportation domains. 
Since GNNs mix the graph and DL computation, the surveyed models demonstrate significant variability, which poses a great challenge for the analysis of their computational characteristics.

\paragraph{Model Decomposition.}
To overcome the diversity challenge, we adopt the recent GNN description framework SAGA-NN~\cite{ngra}, which defines four stages in a GNN layer (\texttt{\underline{S}catter}, \texttt{\underline{A}pplyEdge}, \texttt{\underline{G}ather}, and \texttt{\underline{A}pplyVertex}.).
The current popular GNN libraries DGL~\cite{dgl} and PyG~\cite{pyg} also implicitly follow a similar framework.
We express and decompose the surveyed models into different stages in the \sagaann{} framework. 
We then categorize each stage's operations to mine common patterns and simplify the analysis.

We use GGNN~\cite{GG-NN} in \Fig{fig:saga2-nn-stage} as an example to illustrate the different stages.
In the first \texttt{Scatter} stage, each edge concatenates its source and destination vertex embeddings as $edge_{scatter}$.
In the second \texttt{ApplyEdge} stage, each edge transforms $edge_{scatter}$ with a MLP operation to produce the new edge embedding $edge^{l+1}$.
In the third \texttt{Gather} stage, each vertex first sums the $edge^{l+1}$ of all its inedges (incoming edges) to output $vertex_{gather}$,
and then concatenates it with the vertex's existing embedding $vertex^l$ as the input for the next stage.
In the fourth \texttt{ApplyVertex} stage, each vertex transforms this input with a GRU (Gated Recurrent Unit) operation to the new vertex embedding $vertex^{l+1}$.

\begin{figure}[t]
    \centering
    \vspace{-9pt}
    \includegraphics[trim=2.6cm 2.9cm 2.4cm 2.2cm,clip,width=\linewidth]{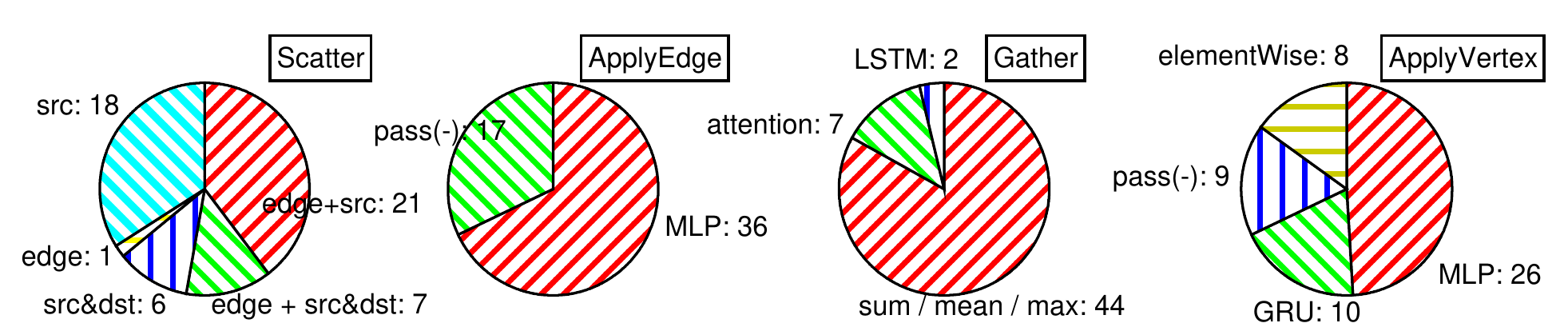}\\
    \vspace{-6pt}
    \caption{The proportion of operations used in each stage of \sagaann{}.}
    \label{fig:algorithm_decompose}
    \vspace{-5pt}
\end{figure}

{\renewcommand{\arraystretch}{1}
\begin{table}[t]
\caption{The \sagaann{} stage breakdown of the selected models.}
\label{tbl:benchmark}
\vspace{-12pt}
\center
\resizebox{\linewidth}{!}{
\begin{tabular}{ c | c | c | c | c  }
\Xhline{0.6pt} 
 \textbf{Model} & 
 \textbf{Scatter} & 
 \textbf{ApplyEdge} & 
 \textbf{Gather} & 
 \textbf{ApplyVertex} \\%
 \Xhline{0.2pt}%
 GCN~\cite{GCN_new} & - & - & sum(in\_edges) & MLP \\%
 GAT~\cite{GAT_new} & src\&dst & MLP & attention(in\_edges) & MLP \\%
 GGNN~\cite{GG-NN} & src\&dst & MLP & sum(all\_edges), vertex & GRU \\%
 R-GCN~\cite{RGCN} & edge+src  & MLP & sum(in\_edges), vertex & sum \\%
 GraphSAGE~\cite{GraphSAGE} & - & - & LSTM(in\_edges), vertex  & MLP \\%
 \Xhline{0.2pt}%
 \textbf{Coverage} & 85\% & 100\% & 100\% & 100\% \\%
 \Xhline{0.6pt}%
\end{tabular}
}
\vspace{-17pt}
\end{table}
}

{\renewcommand{\arraystretch}{1.1}
\begin{table}[b]
\vspace*{-13pt}
\caption{The evaluated datasets for the vertex classification task. }
\label{tbl:dataset}
\vspace{-12pt}
\center
\resizebox{\linewidth}{!}{
\begin{tabular}{ c | c | c | c | c | c | c } 
\Xhline{1.2pt}
\textbf{Name} &  Cora & Citeseer & Pubmed & AIFB & MUTAG& BGS \\\Xhline{0.6pt}
\textbf{Vertex\#} & 2,708 & 3,327&19,717&8,285&23,644&333,845 \\\hline
\textbf{Edge\#} & 5,429& 4,732& 44,338 & 29,043 & 74,227 &916,199 \\\hline
\multirow{2}{*}{\textbf{\makecell{Graph\\Type}}} & \multirow{2}{*}{\makecell{Citation\\Network}}  & \multirow{2}{*}{\makecell{Citation\\Network}} & \multirow{2}{*}{\makecell{Citation\\Network}} & \multirow{2}{*}{\makecell{Semantic\\Network}} & \multirow{2}{*}{\makecell{Molecular\\Structure}} &\multirow{2}{*}{\makecell{Geological\\Graph}} \\ 
 & & & & & & \\ \Xhline{0.6pt}
\end{tabular}
\vspace{-6pt}
}
\end{table}}

\paragraph{Survey Result.}
For the surveyed 53 models, we decompose them into different \sagaann{} stages.
\Fig{fig:algorithm_decompose} summarizes the operation distribution in each stage.
The results show that \textit{although GNNs have a tremendous design space (diversity), there exist a few commonly used operations in each stage.}
GNNs can adopt a mix of source/destination vertex embedding, and edge embedding in the \texttt{Scatter} stage.
For the \texttt{Gather stage}, some models use the simple sum/mean/max operation while others use more complex attention/LSTM (Long short-term memory) operations.  
Some models bypass the \texttt{ApplyEdge} or \texttt{ApplyVertex}, while most models use MLP operations. 
The survey insight suggests that we can cover the large GNN space by carefully selecting a few models.

\paragraph{Studied Models.}
In the end, we select five representative GNN models in Table~\ref{tbl:benchmark}, which cover almost 100\% of the operations used in the 53 surveyed models.
In other words, the 53 models can be viewed as recombination of different stages in the five models.
We study and analyze those models on existing GNN execution library DGL~\cite{dgl} and PyG~\cite{pyg}.
However, the two libraries do not explicitly implement \sagaann{} so we refactor the codes following the \sagaann{} description framework for a more systematic analysis. 
We focus more on DGL analysis because the PyG currently does not support GraphSAGE.
We also select 6 commonly used datasets from the literature, with the number of vertices ranging from 5~K to about 1~M in \Tbl{tbl:dataset}.

\section{GNN Computation Analysis on GPU}
\label{sec:analysis:soft}
 
We study the computational characteristics of the selected GNN models on the contemporary GPU.
We conduct an end-to-end analysis to examine the model difference and the impact of the dataset.
We then analyze the stage-level characteristics in detail, which lets us identify the possible bottleneck and suggest new optimizations.
 
\begin{figure}[t]
    \centering 
    \vspace*{-5pt}
    \hspace*{15pt}
    \includegraphics[width=0.99\linewidth]{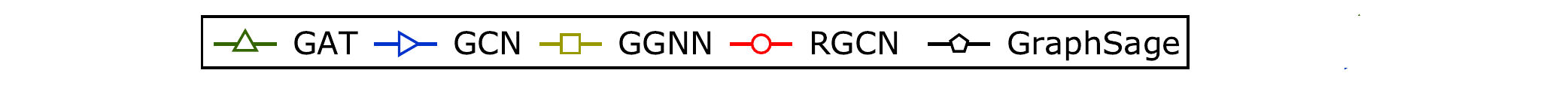}\\%
    \vspace*{-6pt}
    \includegraphics[width=0.33\linewidth]{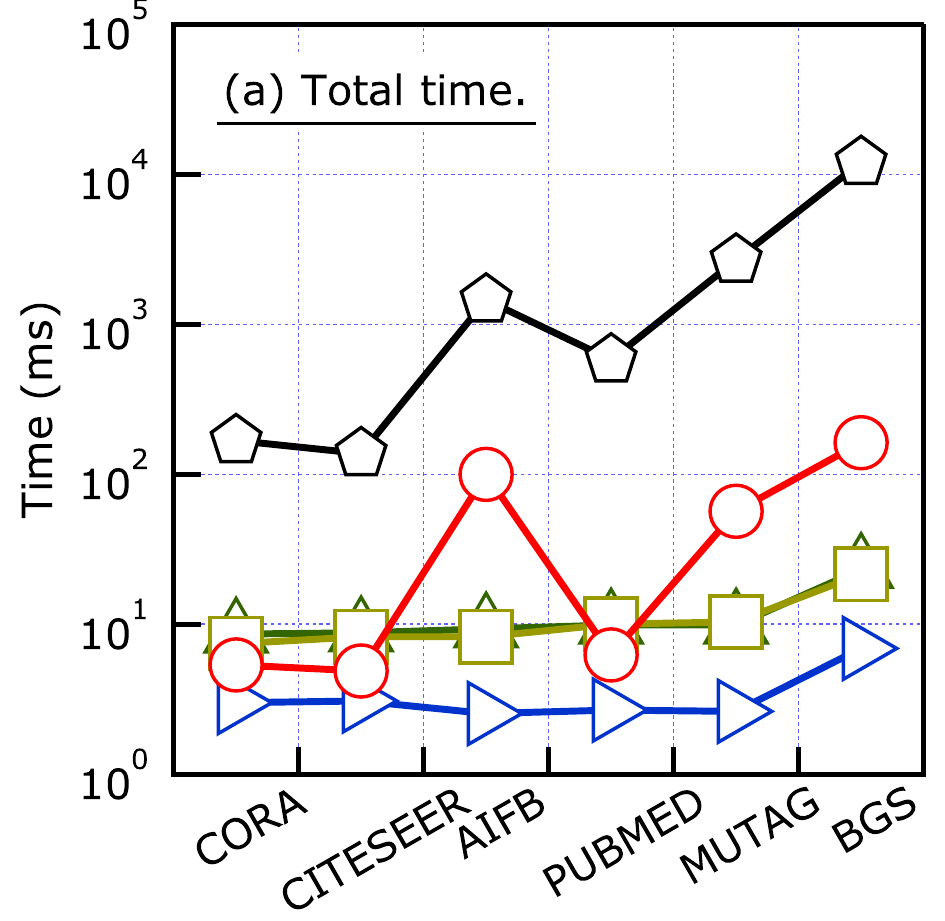}%
    \includegraphics[width=0.33\linewidth]{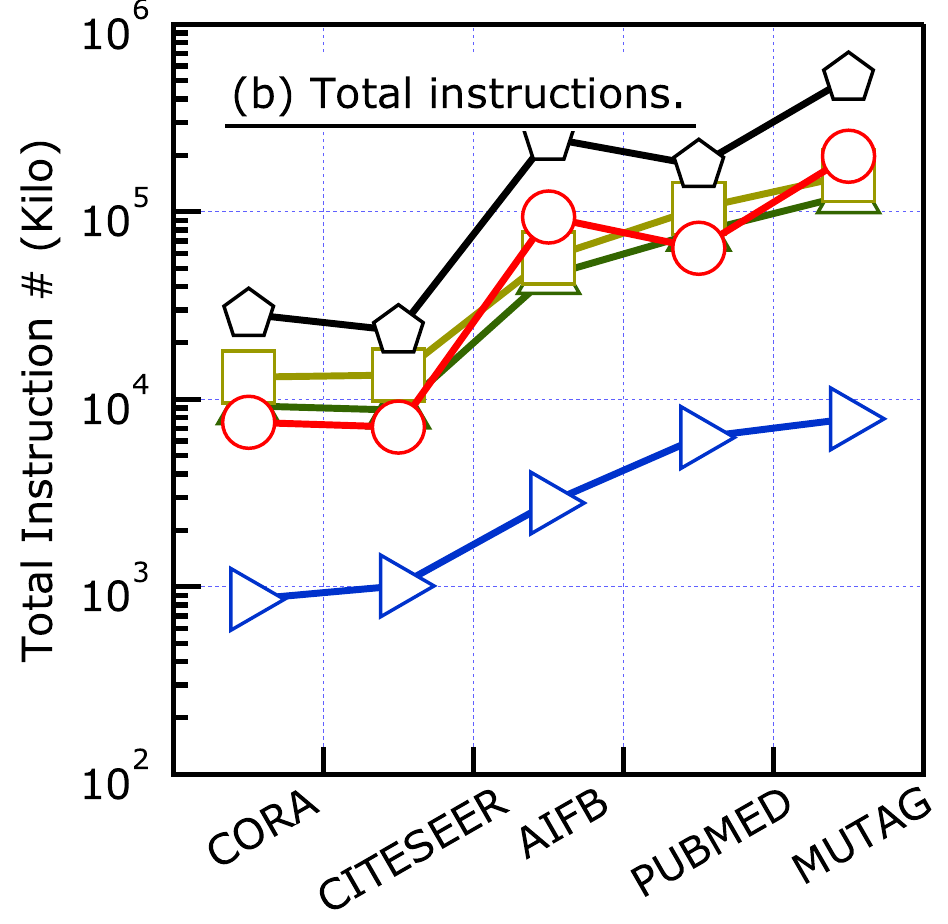}%
    \includegraphics[width=0.33\linewidth]{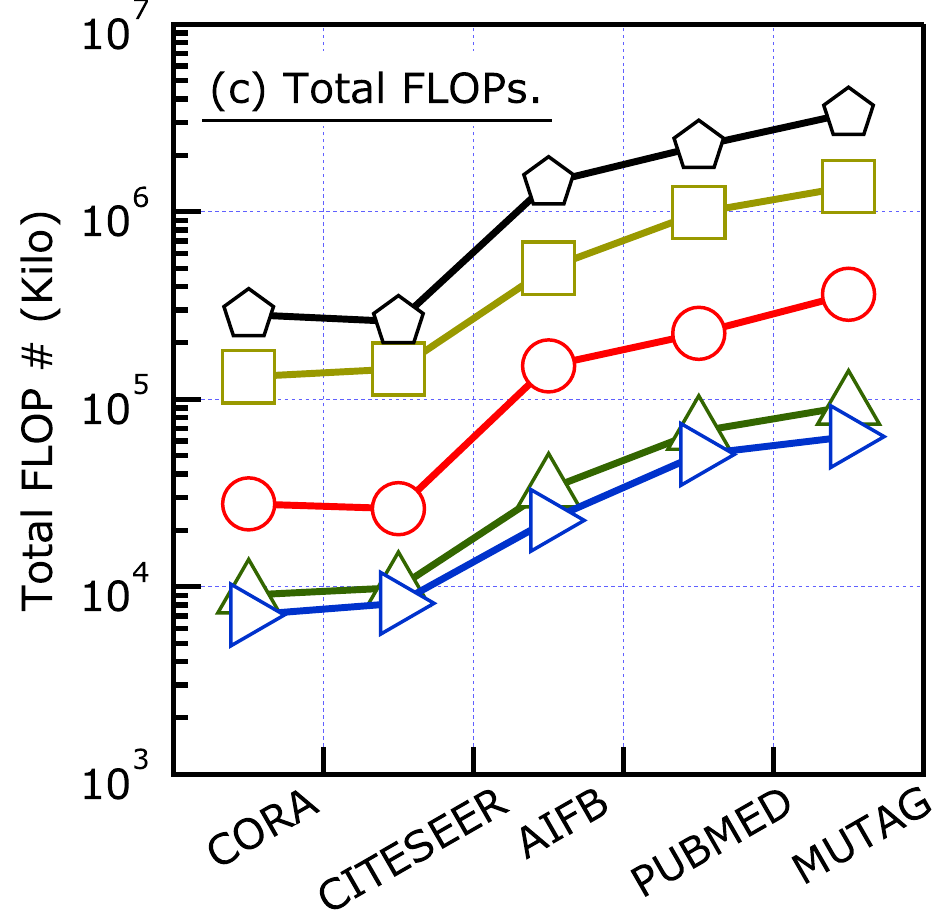}\\%
    \vspace{-5pt}%
    \caption{Execution statistics of selected models with the different datasets.}
    \label{fig:highlevel}
    \vspace{-20pt}
\end{figure}

\subsection{End-to-end Analysis}
\Fig{fig:highlevel} plots the execution statistics of our selected models with the different datasets (\Tbl{tbl:dataset}).
The experiments are carried on a machine with two Intel Xeon 4110 CPUs and one NVIDIA RTX 2080Ti GPU. The machine runs Ubuntu 16.04 with CUDA 10 and cuDNN 7. We use GNN libraries DGL 0.4~\cite{dgl} and PyG 1.3~\cite{pyg}, both with PyTorch 1.4 as backend.
We summarize the key insights as follows.

\paragraph{Execution Time.}
\Fig{fig:highlevel}(a) plots the inference execution time with different graph structures that are ranked by their edge/vertex count.
The inference time for GraphSAGE and RGCN generally increases with graph vertex/edge count except for the AIFB dataset.
The reason for RGCN is that this dataset has additional edge types that lead to more computation in the two models,  while the reason for GraphSAGE is its bottleneck stage \texttt{Gather} that we describe later.
In contrast, the inference time for the rest three models do not vary except the largest graph BGS.
The reason is that the CPU time dominates the execution when the graph is small.

\paragraph{Instruction \& FLOP.}
\Fig{fig:highlevel}(b) and (c) compare the total instructions and FLOPs (floating-point operations), which increase as the graph size increases.
When the graph is too small, the CPU time dominates the entire inference time so the increase of instructions and FLOPs are not reflected in \Fig{fig:highlevel}(a).
Unlike the CNNs whose input is usually the fixed-size image, the size of graphs varies significantly across different domains and problem settings.
As such, the design of GNN acceleration architecture must be aware of the graph size (also embedding vector size) of the targeted problems to balance resource utilization. 

We also observe that GCN is the simplest model with less number of instructions and FLOPs.
The rest four models have similar instruction counts for the same graph but their FLOPs are quite different.
This variability can be attributed to their different model complexity: GraphSAGE is more complicated with the LSTM-based \texttt{Gather} stage while GGNN uses GRU cell in the \texttt{ApplyVertex} stage.
Next, we perform a detailed stage-level analysis for a deeper understanding of these models.

\subsection{Stage-level Analysis}

\begin{figure}[t]
    \centering
    \vspace*{-7pt}%
    \begin{minipage}{.495\linewidth}
    \vspace*{1pt}%
    \centering
    \includegraphics[width=0.99\linewidth]{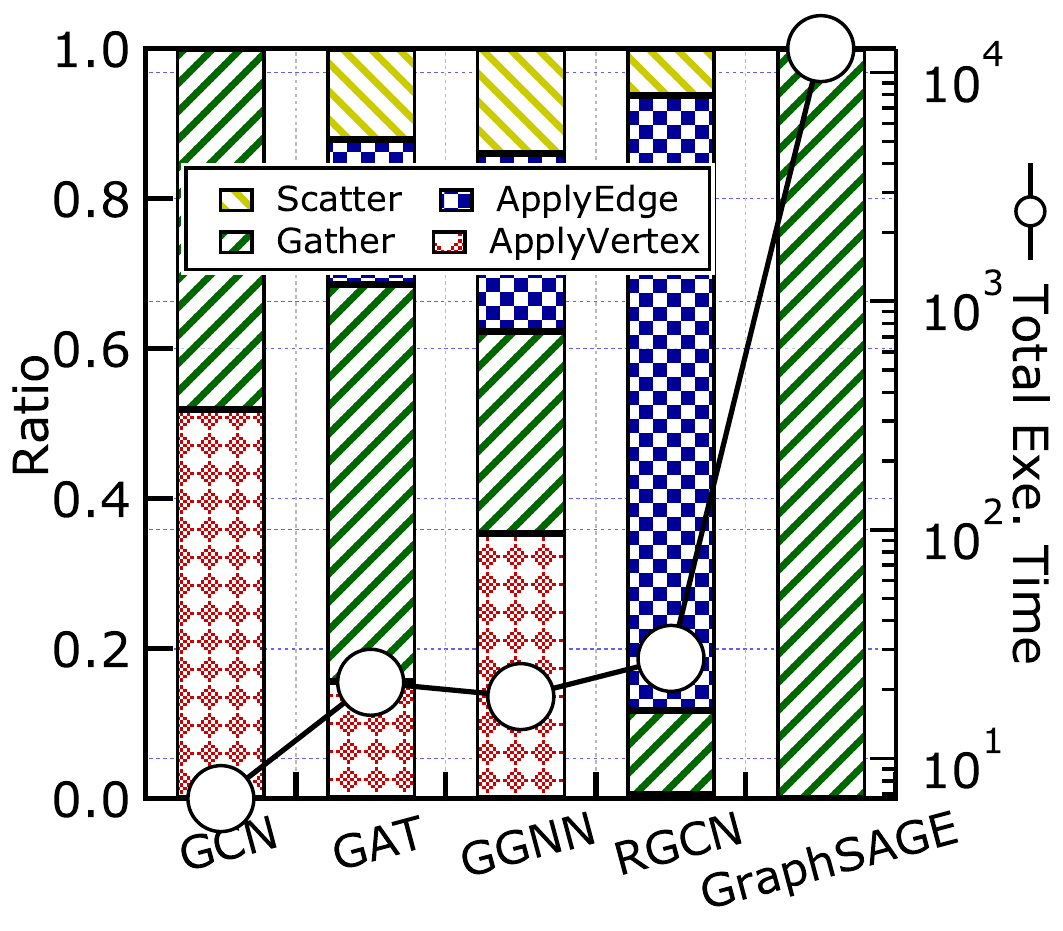}
    \vspace*{-21.5pt}%
    \caption{Stage time breakdown.}
    \label{fig:breakdown:time}
    \end{minipage}\hfill%
    \begin{minipage}{0.495\linewidth}
    \centering
    \vspace*{1pt}
    \includegraphics[width=0.99\linewidth]{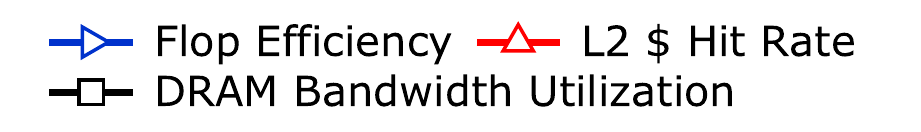}\\%
    \vspace*{-3pt}
    \includegraphics[width=0.495\linewidth]{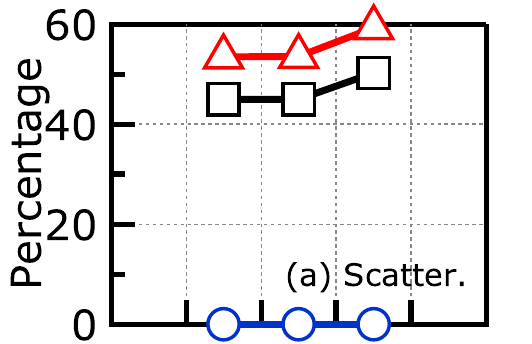}%
    \includegraphics[width=0.495\linewidth]{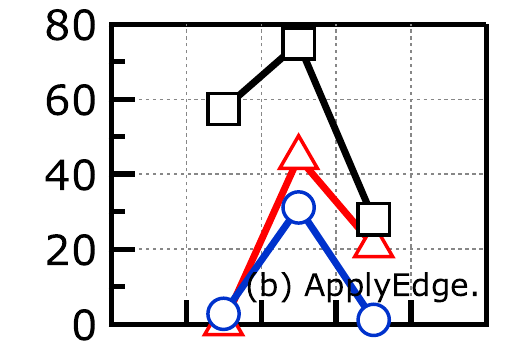}\\%
    \vspace*{-1.5pt}
    \includegraphics[width=0.495\linewidth]{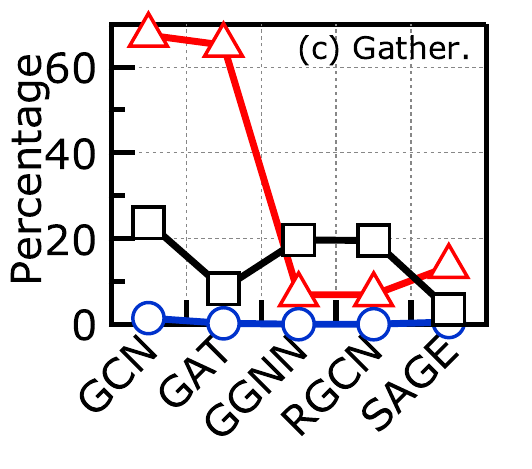}%
    \includegraphics[width=0.495\linewidth]{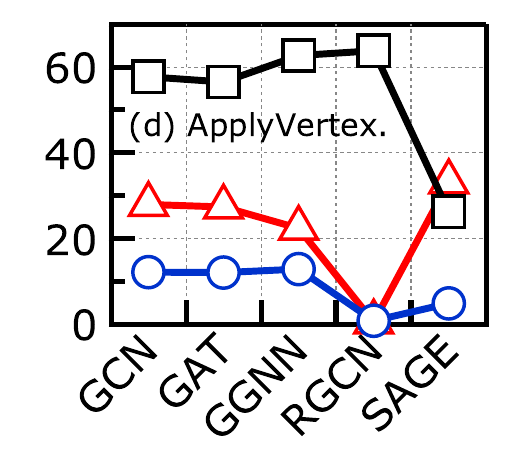}%
    \vspace*{-9pt}%
    \caption{Stage execution statistics.}
    \label{fig:breakdown:stats}
    \end{minipage}
    \vspace*{-19pt}%
\end{figure}

\Fig{fig:breakdown:time} shows the stage-level time breakdown and \Fig{fig:breakdown:stats} shows each stage's execution statistics. Both results are obtained on the largest dataset BGS.
The stage time distribution varies greatly among models, which confirms the diversity of our selected models, and also suggests that there is no fixed hotspot in GNNs. Therefore, we must equally and jointly consider all the stages. 
We present our detailed stage-by-stage analysis as follows.

\paragraph{Scatter.}
This stage prepares data for its following edge transformation stage \texttt{ApplyEdge} so that it only involves data movement.
As such, \Fig{fig:breakdown:stats}(a) shows that this stage has no floating-point operation and intensively uses DRAM bandwidth (note that GCN and GraphSAGE bypass this stage).
Through the kernel trace profiling (not shown owing to space limits), we find that this stage uses the CUDA kernel \texttt{indexSelectLargeIndex} to implement the data movement, which puts together the embedding vector of each edge and the embedding vector of its two connected vertices into a new edge embedding vector (\Fig{fig:saga2-nn-stage}). 
This stage also has a high L2 cache hit rate while graph processing is well known as low locality. 
The reason is that GNNs use a typical size of 32 or more for vertex/edge embedding while graph processing like BFS/PageRank uses one.

\paragraph{ApplyEdge.}
This stage performs edge embedding matrix transformation with MLPs.
In general, it is possible to batch all edges for a parallel process so that this stage should have high computation efficiency.
However, only GGNN has high FLOP efficiency while both GAT and RGCN have low-efficiency values.
The size of edge embedding in GAT and GGNN is 1 (as attention value) and 32, respectively.
As such, the \texttt{ApplyEdge} stage can be implemented by GEMV (matrix-vector multiplication) and GEMM in GAT and GGNN, respectively.
This explains why the FLOP efficiency in GGNN is higher than the GAT.
Meanwhile, the RGCN assigns another edge type attribute to different edges.
As a result, its \texttt{ApplyEdge} stage first needs to select edges with the same type and then batches them to GEMM operation, which leads to the overall low FLOP efficiency.

\begin{figure}[t]
    \centering
    \vspace{-6pt}%
    \includegraphics[width=0.98\linewidth]{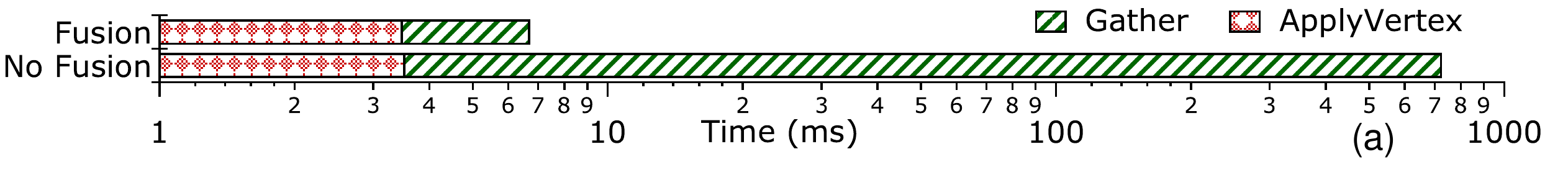}~~~\\%
    \vspace{-2pt}\hspace{-3pt}%
    \includegraphics[width=0.90\linewidth]{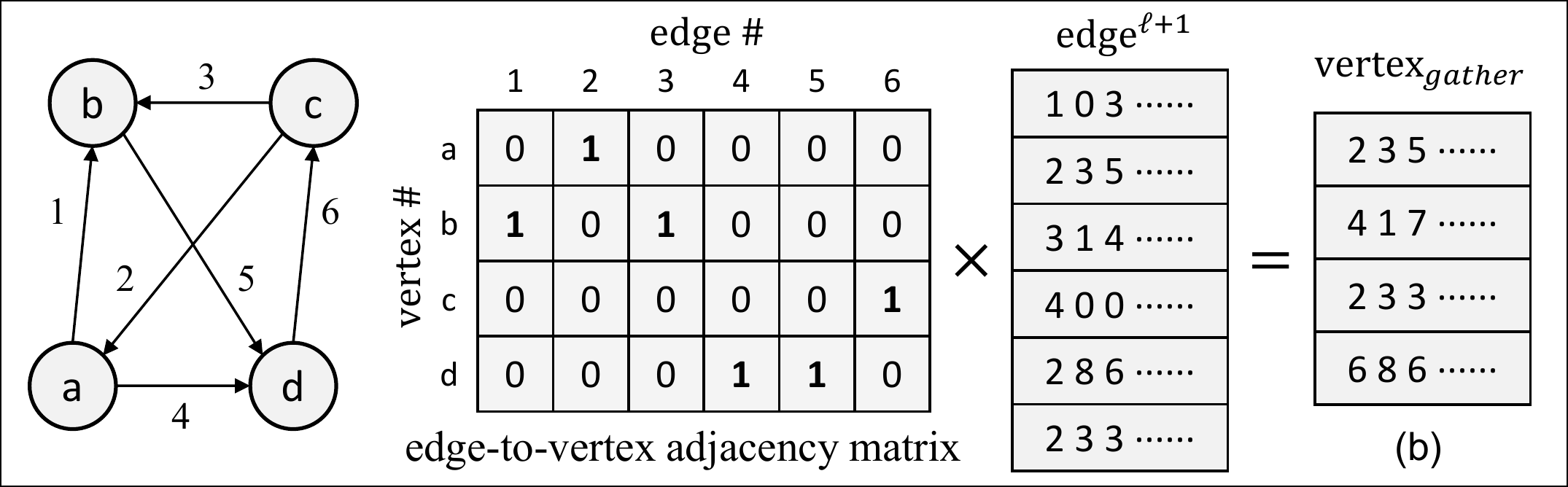}\\%
    \vspace{-5pt}%
    \caption{\scriptsize (a) GCN time with/without fusion. (b) The fused \texttt{Gather} can be implemented by multiplying the graph adjacency matrix and edge embedding matrix.}
    \vspace{-18pt}
    \label{fig:combination}
\end{figure}
 
\noindent\textbf{Gather.}
This stage gathers edge embeddings for the following vertex transformation stage \texttt{ApplyVertex}.
Since different vertices have different edge counts, this stage uses a \texttt{Reduction} function that reduces the varying edge count to a fixed size embedding. 
Through profiling, we find that the \texttt{Gather} stage uses an important fusion optimization.
To emphasize its importance, we run the GCN model with and without the fusion that DGL exposes. 
\Fig{fig:combination}-a shows that the stage fusion leads to about $200 \times$ speedup.

We use an example with the \texttt{sum} reduction function to illustrate how the fusion works.
Assume an edge-to-vertex adjacency matrix (\Fig{fig:combination}-b middle), where a nonzero element represents an inedge (column) of the corresponding vertex (row).
With the sum reduction function, the \texttt{Gather} stage calculates the new embedding of each vertex by summing the embeddings of all inedges.
For each vertex, the summation can be implemented by multiplying its corresponding row in the adjacency matrix and the entire edge embedding matrix, where each row is an edge embedding vector.
As such,  the fused operation equals the multiplication between the adjacency matrix and the embedding matrix.
Through the kernel trace profiling, we find that \emph{this multiplication uses sparse GEMM for great performance and efficiency as the adjacency matrix is highly sparse.}

\begin{figure}[t]
    \centering
    \vspace{-7pt}%
    \hspace{-3pt}%
    \includegraphics[scale=0.5]{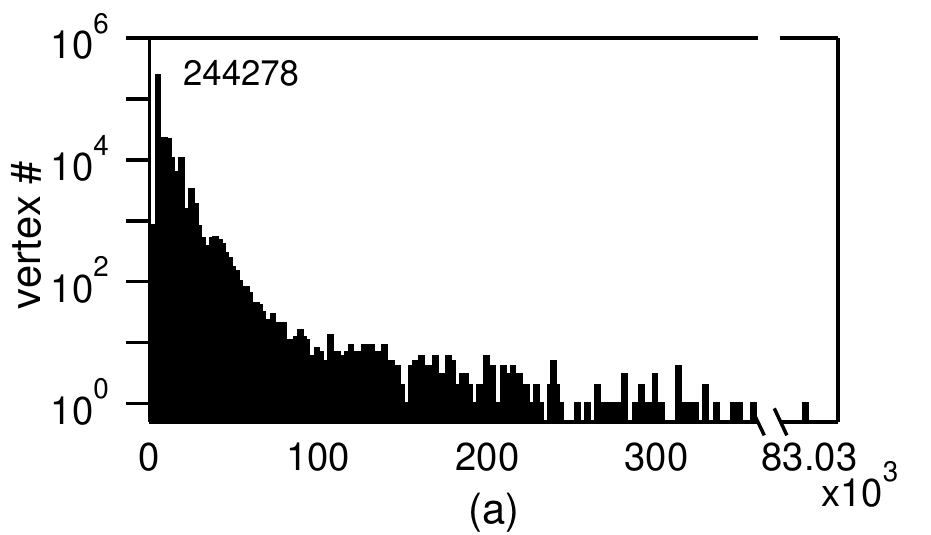}~%
    \hspace{-6pt}%
    \includegraphics[scale=0.5]{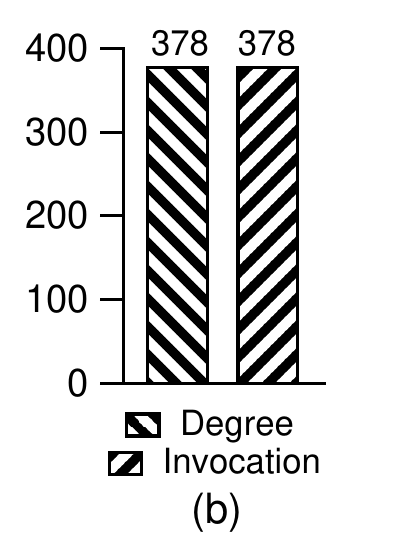}%
    \includegraphics[height=0.3\linewidth]{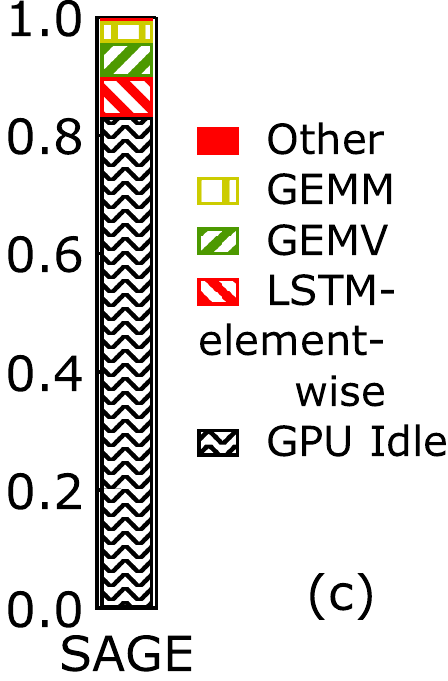}\\
    \vspace{-8pt}
    \caption{\scriptsize (a) Degree distributions of BGS. (b) The number of degrees in the graph and invocations in the \texttt{Gather} stage. (c) Time distribution of kernels in the \texttt{Gather} stage of GraphSAGE.}
    \vspace{-20pt}
    \label{fig:degree}
\end{figure}

The exception is GraphSAGE which adopts \texttt{LSTM} as the reduction function, which treats the inedges of a vertex as a sequence and outputs a new vector.
The \texttt{LSTM} requires a serialized input of a vertex's inedges, which cannot leverage the fusion optimization and becomes the bottleneck (\Fig{fig:breakdown:time}). 
We find that DGL implements the LSTM-based \texttt{Gather} in a \textit{degree traversal} technique. It batches the vertices with the same degree for parallel computation on the GPU. 
\Fig{fig:degree}-b shows that the number of kernel invocations equals the number of vertex degrees in the graph.

The degree traversal approach causes a severe GPU under-utilization as the idle time dominates in the stage (\Fig{fig:degree}-c).
\Fig{fig:degree}-a shows the degree distribution in the BGS dataset, which obeys the power-law distribution: the number of vertices with the same degree decreases exponentially when the degree increases.
As such, this approach is close to the sequential vertex-by-vertex computation for large degrees.
Through profiling, we find that the interval time between two adjacent vertex computations can be 20$\times$ larger than the invoked kernel duration.
As a result, the \texttt{Gather} becomes the major bottleneck in GraphSAGE.

\paragraph{ApplyVertex.}
This stage usually performs vertex embedding matrix transformation with MLPs, which is similar to the \texttt{ApplyEdge} stage.
All vertices can be batched so that this stage can be implemented by GEMM, which leads to the highest FLOP efficiency among all stages (\Fig{fig:breakdown:stats}d).
The only exception of RGCN is due to the simple \texttt{sum} defined for the stage (\Tbl{tbl:benchmark}).
On the other hand, \texttt{ApplyEdge} and \texttt{ApplyVertex} often play a key role in the model accuracy, and more and more GNN models are being developed with complex functions in those two stages to capture more information on the vertex, edge, and graph.

{\renewcommand{\arraystretch}{1.1}
\begin{table}[b]
\vspace*{-15pt}
\caption{The characteristics of different GNN computation stages.}\label{tbl:regularity}
\vspace{-6pt}
\centering
\resizebox{\linewidth}{!}{
\begin{tabular}{c | c | c} 
\Xhline{0.6pt} 
\textbf{Stage} & \textbf{Description} & \textbf{Kernel} \\
 \Xhline{0.6pt}
 Scatter & Vertex/edge embedding movement & IndexSelection\\
\Xhline{0.2pt}
 ApplyEdge & DL-based edge embedding transformation& GEMM/GEMV\\
\Xhline{0.2pt}
Gather (fused) & Edge  embedding reduction & Sparse GEMM \\
\Xhline{0.2pt}
Gather & Edge embedding movement &  IndexSelection \\
(non-fused)& Complex reduction (e.g., LSTM) & GEMM/GEMV \\                       
\Xhline{0.2pt}
 ApplyVertex & DL based vertex embedding transformation & GEMM/GEMV\\
\Xhline{0.6pt}
\end{tabular}}
\vspace*{-0pt}
\end{table}}

\paragraph{Library.}
We observe a similar result on PyG, especially for \texttt{ApplyEdge/ApplyVertex} that both leverage PyTorch's existing features for efficient computation.
Their greatest difference lies in \texttt{Scatter} and \texttt{Gather} stages.
For example, PyG lets users customize the data movement in \texttt{Scatter} while DGL moves all data (embeddings of the edge and its vertices) by default.
As such, PyG performs better when part of the data is fetched (e.g., RGCN).
Moreover, the PyG does not support the LSTM-based \texttt{Gather} stage.

\paragraph{Summary.} 
Table~\ref{tbl:regularity} summarizes the characteristics of each stage. 
It shows that although GNNs have a large design space, the stage-level characteristics are relatively stable across different models.
In other words, our stage-level characterization can lay the foundation for future hardware/software acceleration of GNNs.

\section{Implications for Hardware Acceleration}
\label{sec:analysis:arch}

We now study the feasibility and challenges for the GNN accelerator, which can further improve the performance or energy efficiency of GNNs, on the basis of our previous analysis.
Instead of building an accelerator from scratch, we take the existing DL accelerator TPU, and study its performance of running the GNN models.
Based on the detailed stage-level analysis, we shed light on the efficient GNN accelerator design.
We use TensorFlow 1.12 on top of one Google Cloud TPU v2 for this part of the experiments.

\begin{figure}[t]
    \centering 
    \vspace{-6pt}%
    \includegraphics[width=0.98\linewidth]{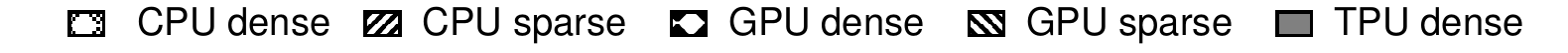}%
    \vspace{-4pt}%
    \\%
    \hspace{-4pt}%
    \includegraphics[width=0.47\linewidth]{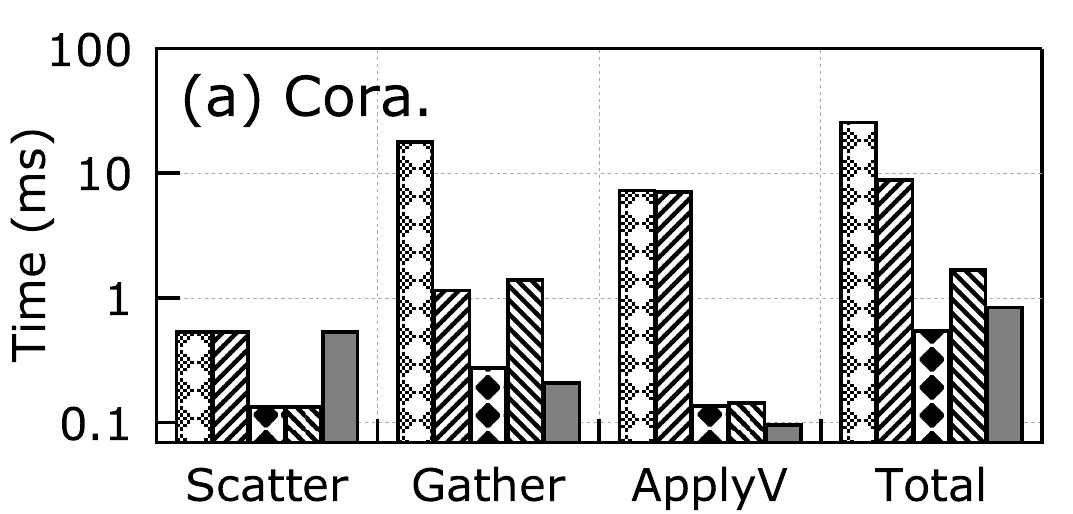}%
    \hspace*{10pt}
    \includegraphics[width=0.47\linewidth]{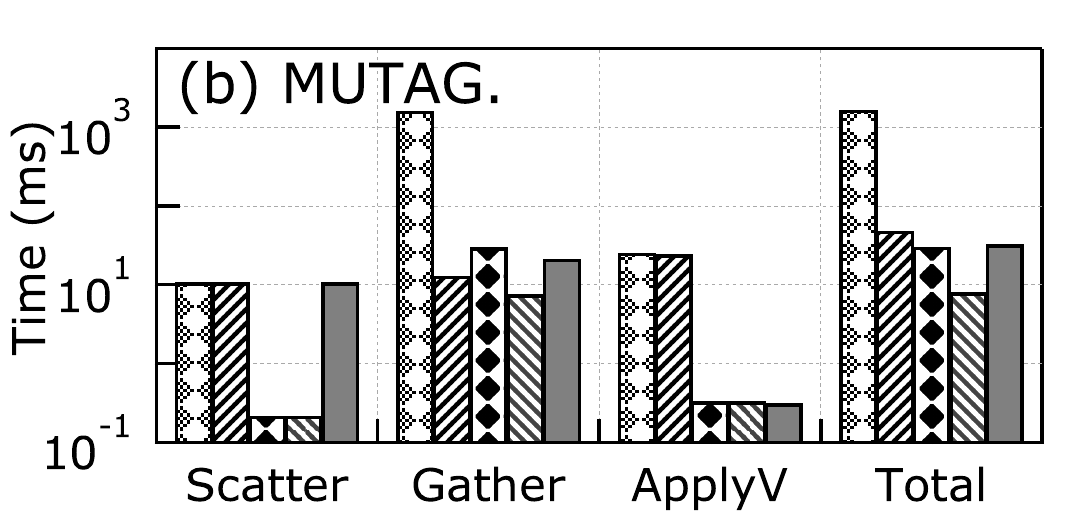}%
    \\%
    \vspace{-6pt}
    \caption{The inference time (ms) on difference architectures.}
    \label{fig:tpu}
    \vspace{-17pt}
\end{figure}

\paragraph{Projection Methodology.}
Because both DGL and PyG only support CPU and GPU platforms, we adopt a micro-benchmark based methodology to project the performance of running a typical GNN model on the TPU.
We extract the parameters for the regular computation and run them on the TPU.
For the irregular data movement, we use the time of our local CPU due to the lack of native TPU support.
\Fig{fig:tpu} shows the result of two datasets.
We report both the dense and sparse performance on the CPU/GPU, and report only the dense performance of TPU because it does not support the sparse GEMM.
 
\paragraph{Result Analysis.}
We show a counter-intuitive result in \Fig{fig:tpu}. The TPU does not achieve the best performance on both datasets (Total bar), although it does achieve the best dense GEMM performance (ApplyV bar).
On the small dataset Cora, TPU performs worse than the GPU because the TPU has to rely on CPU for the data movement.
On the large dataset MUTAG, the GPU outperforms TPU with the sparse stage fusion due to the higher graph sparsity.

\paragraph{Forward Looking.} 
In summary, we think it is important for GNN accelerators to support both sparse and dense matrix operations for efficient GNN acceleration. Efficient data movement for the graph structure is also indispensable to avoid unnecessary data copy between processors. 
As such, an ideal GNN accelerator in our outlook consists of three key components: data movement component, dense operation component, and sparse operation component.
Meanwhile, the current execution paradigm for GNN is stage-by-stage and layer-by-layer. 
We believe that there is a huge design space to break this sequential paradigm by, for example, fine-grained pipelining the different stages.
We leave those for our future work.

\section{Conclusion}\label{sec:conclusion}

We systematically study the computation characteristics of graph neural networks. 
We first construct a representative GNN benchmark based on the extensive model review and a general GNN description framework. 
We then analyze their computational efficiency and microarchitectural characteristics on the existing GPU architecture.
Our analysis suggests that the GNN is a unique workload with the mixed features from graph analytics and DL computation, which warrants more future research.

\ifCLASSOPTIONcompsoc
  \section*{Acknowledgments}
\else
  \section*{Acknowledgment}
\fi

We thank the anonymous reviewers for their constructive feedback.
This work was supported by National Key R\&D Program of China (2018YFB1305900), the National Natural Science Foundation of China Grant (61702328, 61832006, and 61972247), Microsoft Research Asia Collaborative Research Grant. Any opinions, findings, and conclusions in this paper are those of the authors only.

\ifCLASSOPTIONcaptionsoff
  \newpage
\fi




{
\bibliographystyle{IEEEtran}
\bibliography{ref}
}


\end{document}